\documentclass{aastex631}

\usepackage{graphicx}
\usepackage{color}
\usepackage{epsfig}
\usepackage{epsf}
\usepackage{epstopdf}
\usepackage{subfigure}
\usepackage{upgreek}
\usepackage{amssymb}
\usepackage{amsmath}
\usepackage{natbib}
\usepackage{graphicx}
\usepackage{color}
\usepackage{tabularx}
\usepackage{rotating}
\usepackage{multirow}
\usepackage{subfigure}
\usepackage{ulem}
\usepackage{longtable}
\begin{document}

%
%

\title{Multi-mode $\delta$ Sct stars from the Zwicky Transient Facility Survey}

\author[0000-0002-0786-7307]{Qi Jia}
\affiliation{School of Physics and Astronomy, China West Normal University, Nanchong 637009, China}
\affiliation{CAS Key Laboratory of Optical Astronomy, National Astronomical Observatories, Chinese Academy of Sciences, Beijing 100101,  China}

\author[0000-0001-7084-0484]{Xiaodian chen}
\affiliation{CAS Key Laboratory of Optical Astronomy, National Astronomical Observatories, Chinese Academy of Sciences, Beijing 100101,  China}
\affiliation{School of Physics and Astronomy, China West Normal University, Nanchong 637009, China}
\affiliation{School of Astronomy and Space Science, University of the Chinese Academy of Sciences, Beijing, 100049, China}
\affiliation{Institute for Frontiers in Astronomy and Astrophysics, Beijing Normal University,  Beijing 102206, China}

\author[0000-0003-4489-9794]{Shu Wang}
\affiliation{CAS Key Laboratory of Optical Astronomy, National Astronomical Observatories, Chinese Academy of Sciences, Beijing 100101,  China}
\affiliation{School of Physics and Astronomy, China West Normal University, Nanchong 637009, China}

\author{Licai Deng}
\affiliation{CAS Key Laboratory of Optical Astronomy, National Astronomical Observatories, Chinese Academy of Sciences, Beijing 100101,  China}
\affiliation{School of Physics and Astronomy, China West Normal University, Nanchong 637009, China}
\affiliation{School of Astronomy and Space Science, University of the Chinese Academy of Sciences, Beijing, 100049, China}

\author{Yangping Luo}
\affiliation{School of Physics and Astronomy, China West Normal University, Nanchong 637009, China}

\author{Qingquan Jiang}
\affiliation{School of Physics and Astronomy, China West Normal University, Nanchong 637009, China}

\correspondingauthor{Xiaodian Chen}
\email{chenxiaodian@nao.cas.cn}

\begin{abstract}

We obtain the largest catalog of multi-mode $\delta$ Sct stars in the northern sky to date using the Zwicky Transient Facility (ZTF) Data Release 20 (DR20). The catalog includes 2254 objects, of which 2181 are new to our study. Among these multi-mode $\delta$ Sct stars, 2142 objects are double-mode $\delta$ Sct, while 109 objects are triple-mode $\delta$ Sct and 3 are quadruple-mode $\delta$ Sct. By analyzing the light curves in the $r$ and $g$ bands of the ZTF, we determine the basic parameters of multi-mode $\delta$ Sct stars, including the periods and amplitudes. Periods are checked by comparison with the OGLE catalog of double-mode $\delta$ Sct stars. On the Petersen diagram, multi-mode $\delta$ Sct stars form six sequences. We find that in Galactic coordinates, the periods of 1O/F double-mode $\delta$ Sct stars at high latitudes are shorter than those of 1O/F double-mode $\delta$ Sct stars in the disk, due to metallicity variations. In the future, our catalog can be used to establish the period--luminosity relation and the period--metallicity relation of double-mode $\delta$ Sct stars, and to study the Galactic structure.

\end{abstract}
\keywords{ Periodic variable stars (1213) ; Pulsating variable stars (1307) ; Delta Scuti variable stars (370) ; Stellar oscillations (1617) ; Distance indicators (394) ; Astronomy Databases (83) }
\section{Introduction}\label{sec:intro}

Delta Scuti ($\delta$ Sct) stars are pulsating variable stars with periodic variations in brightness \citep{2000ASPC..210....3B}. These stars belong to the Population I stars, while their Population II counterparts are metal poor, classified as SX Phoenecis stars (SX Phe). $\delta$ Sct stars have spectral types between A0 and F5, effective temperatures in the range between 6500K and 8500K. The masses of this class of stars range from about $1.5 \leq M \leq 2.5 M_\odot$ \citep{2017ampm.book.....B}. $\delta$ Sct stars exhibit a wide range of pulsation periods, usually between 0.02 and 0.3 days. Their peak-to-peak amplitudes in visible band ($V$-band) can reach approximately 0.9 mag \citep{2000ASPC..210..373M,2011A&A...534A.125U,2011MNRAS.417..591B,2013AJ....145..132C,2017ampm.book.....B}. In the Hertzsprung-Russell (HR) diagram, $\delta$ Sct stars are located in the region of lower part of classical Cepheid instability strip and with luminosities ranging from the zero-age main sequence (ZAMS) to about 2 mag above the main sequence. These variable stars may also appear in the pre-main sequence (pre-MS) or post-main-sequence (post-MS) phases \citep{2001A&A...366..178R,2004A&A...414L..17D,2013AJ....145..132C}. Such positional variations are due to changes in their evolutionary state and internal structure.

The pulsating phenomenon exhibited by $\delta$ Sct stars in their unique evolutionary state provides a rare opportunity to investigate the structure and evolution of these interesting objects \citep{1998A&A...332..958B,2019MNRAS.485.2380M}. It has been observed that most of $\delta$ Sct stars have amplitude modulations accompanied by period variations, representing different evolutionary stages  \citep{2015EPJWC.10106013B, 2016PhDT.......370B, 2016MNRAS.460.1970B}. Thus, the identification of modes of $\delta$ Sct stars is an important research focus in asteroseismology \citep{1999MNRAS.302..349B,2022ApJ...936...48Y,2022AJ....164..218L}. Mode identification is performed by analyzing the light curves (LCs) of $\delta$ Sct stars \citep{1993A&A...271..482B}. Most $\delta$ Sct stars show non-radial mode while a few of high-amplitude $\delta$ Sct stars show radial pressure modes (p mode) which is excited by the $\kappa$ mechanics \citep{2000ASPC..210....3B}. 
The $\delta$ Sct stars usually pulsate simultaneously in more than one mode \citep{2011A&A...534A.125U}. Double-mode $\delta$ Sct star is a subtype of $\delta$ Sct stars that mainly exhibits two radial pulsation periods, namely the fundamental (F) mode, first-overtone (1O) mode. For most of double-mode $\delta$ Sct star, the primary period is longer than the secondary period. There are also a number of double-mode $\delta$ Sct stars whose secondary period is in a low-amplitude non-radial mode \citep{2016MNRAS.460.1970B}. It is also possible to have three radial modes or four radial modes at the same time, which we call triple-mode and quadruple-mode $\delta$ Sct \citep{1973A&A....23..221B,1979PASP...91....5B,1996A&A...312..463P,1999A&A...341..151B,1999MNRAS.302..349B,2000ASPC..210....3B,2000A&AS..144..469R,2005A&A...438..653D}. By plotting the relation between the period ratio and the period also known as the Petersen diagram \citep{1973A&A....27...89P}, it is possible to distinguish different double-mode $\delta$ Sct stars and to constrain the evolutionary state of double-mode $\delta$ Sct stars. \cite{2021AJ....162...48L, 2022AJ....164..218L} adopted Kepler data to analyze the periods of two double-mode $\delta$ Sct stars.

The period--luminosity relation (PLR) of $\delta$ Sct stars is an important tool for measuring cosmic distances. Distances can be estimated by comparing the apparent magnitude with the absolute magnitude determined by the PLR. Recently, \cite{2019MNRAS.486.4348Z} and \cite{2022MNRAS.516.2080B} used parallaxes from Gaia Data Release 2 and 3 (DR2 and DR3) to derive more accurate PLRs for $\delta$ Sct stars. They found that $\delta$ Sct stars with higher amplitudes tend to be closer to the ridges of the PLR, while those with lower amplitudes show a more dispersed distribution. Another study by \cite{2020MNRAS.493.4186J} used Gaia DR2 parallaxes to determine the PLRs of $\delta$ Sct stars in ASAS-SN and found a strong correlation between the periods of $\delta$ Sct stars and their metallicities. This correlation suggests that the period of $\delta$ Sct stars increases as the Galactocentric radius increases, indicating a gradient in metallicity. Moreover, \cite{2022ApJ...940L..25M} and \cite{2023A&A...674A..36G} also reported a segmented PLR for extragalactic $\delta$ Sct stars. But the fragmentation PLR was denied by the recent PLRs established by \cite{2023arXiv230915147S} from the OGLE LMC sample.

Over the past two decades, high-cadence optical telescopes have been used to monitor stars. The ground-based telescopes, such as Zwicky Transient Facility\footnote{\url{https://www.ztf.caltech.edu/index. html}} \citep[ZTF,][]{2019PASP..131g8001G, 2019PASP..131a8002B} and the Optical Gravitational Lensing Experiment\footnote{\url{http://ogle.astrouw.edu.pl}} \citep[OGLE,][]{2015AcA....65....1U}, and the space telescope missions, such as Kepler\footnote{\url{https://archive.stsci.edu/missions-and-data/kepler}} \citep{2010Sci...327..977B, 2010ApJ...725.1226S} and  Gaia\footnote{\url{https://www.cosmos.esa.int/web/gaia}} \citep{2016A&A...595A...1G} have provided a large amount of high-precision and long-term photometric data for the study of $\delta$ Sct stars. \cite{2020ApJS..249...18C} identified 16,709 $\delta$ Sct stars using the ZTF survey, and \cite{2020MNRAS.493.4186J} found 8,400 $\delta$ Sct stars in All-Sky Automated Survey for SuperNovae (ASAS-SN). \cite{2020AcA....70..241P} and \cite{2021AcA....71..189S} published a catalog of over 24,000 $\delta$ Sct stars based on the OGLE data. \cite{2011MNRAS.417..591B} analyzed 1568 $\delta$ Sct stars using the Kepler data. Gaia detects about 100,000 OBAF-type pulsating stars in the main sequence, of which more than 14,000 are $\delta$ Sct stars \citep{2023A&A...674A..36G}.

With the rapid increase of $\delta$ Sct stars, it is important to identify multi-mode $\delta$ Sct stars. \cite{2023NatAs.tmp..126C} found that the PLR of the double-mode RR Lyrae is not affected by the metallicity, and it is still worth investigating whether this conclusion applies to $\delta$ Sct stars. In this work, we search for multi-mode $\delta$ Sct stars based on ZTF DR20. Our data processing and methods are described in Section \ref{sec:dataandmethod}. Section \ref{sec:result} shows the result of multi-mode $\delta$ Sct stars in ZTF. In Section \ref{sec:statistics}, we discuss the statistic properties of 1O/F double-mode $\delta$ Sct stars. Section \ref{sec:conclusion} summarizes this work.

\section{Data and Method}\label{sec:dataandmethod}
 \subsection{ZTF data}\label{sec:data}	

ZTF is a 48-inch Samuel Oschin Schmidt telescope at the Palomar Observatory with a field of view of 47 deg$^2$. The main scientific goals are to explore transient phenomena in the universe, stellar variability and solar system science \citep{2019PASP..131g8001G, 2019PASP..131a8003M}. The ZTF public surveys are divided into two phases, the first of which runs from March 2018 through September 2020. It obtains photometric measurements in the $g$ and $r$ bands with a three-night cadence in the northern sky and a one-night cadence in the Galactic plane. The uniform exposure time for each observation is 30 seconds. The second phase, starting in December 2020, devotes 50\% of the available observation time to a two-night cadence photometry in the $g$ and $r$ bands of the northern sky \citep{2019PASP..131a8002B}.

Based on ZTF DR2 (time span of $\sim$ 470 days), \cite{2020ApJS..249...18C} published a catalog with $\sim$ 780, 000 periodic variable stars. We selected 16709 $\delta$ Sct stars from this catalog, and cross-matched them with the ZTF DR20\footnote{\url{https://www.ztf.caltech.edu/ztf-public-releases.html}} with a radius of 1'' to obtain the $g$ and $r$ band photometry. The mean number of photometry in the $g$ and $r$ band is around 300 and 500, respectively, which is very suitable for discovering multi-mode $\delta$ Sct stars. We searched for multi-mode $\delta$ Sct stars based on $g, r$ band LCs.
\begin{figure*}
	\begin{center}
		\includegraphics[width=1.\linewidth]{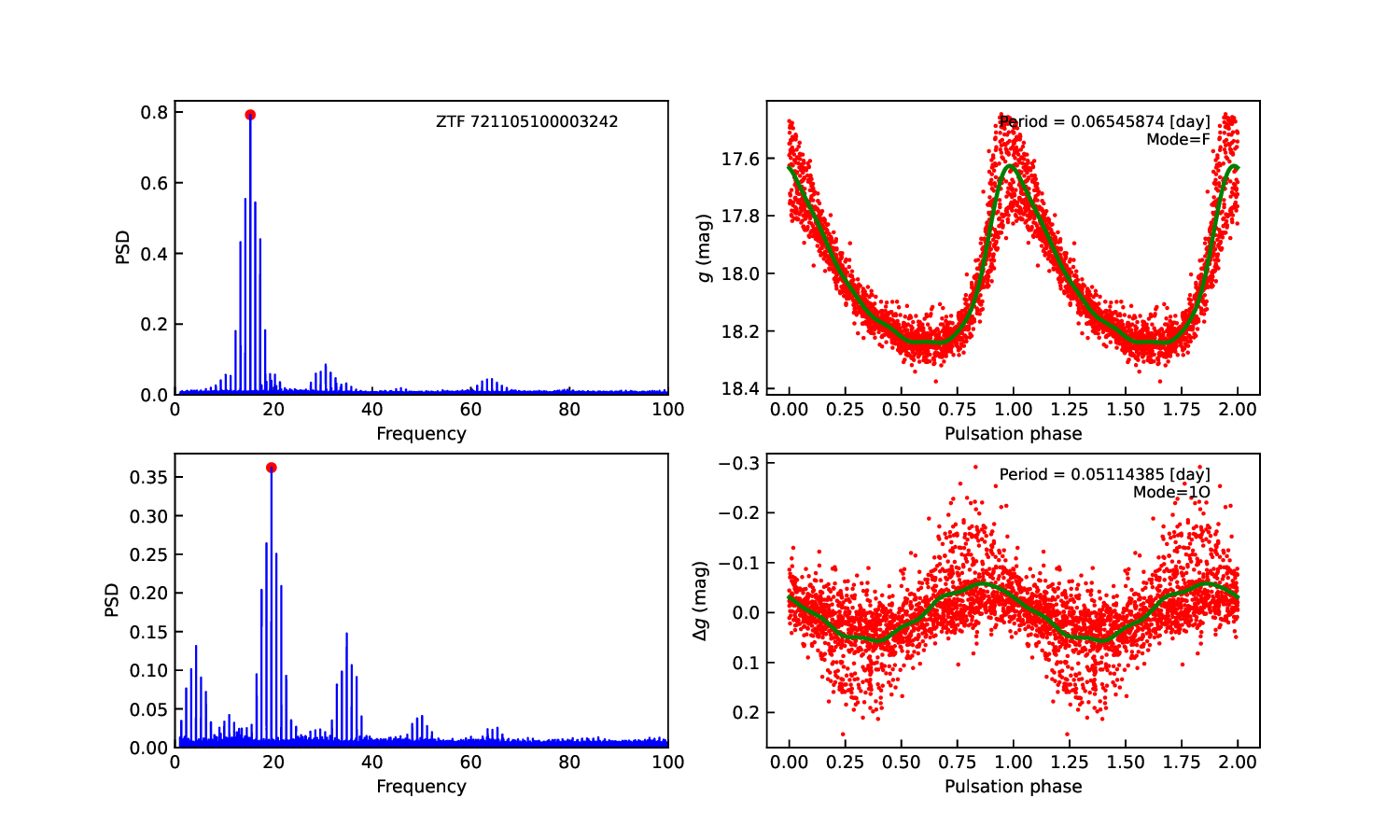}
		\caption{Example diagram of 1O/F double-mode $\delta$ Sct stars for $g$-band LC and PSD. Top left: PSD of the original data, the red filled circle shows the maximum PSD. Top right: red dots are original LC folded by the first period, the green line is the best Fourier fit. Bottom left: PSD of residual LC after prewhitening, the red filled circle shows the maximum PSD. Bottom right: red dots are residual LC after prewhitening folded by the second period, the green line is the best Fourier fit to the residual LC.}
		\label{LCs1}
	\end{center}
\end{figure*}

\subsection{Method}\label{sec:method}	
We analyzed the periodic signals from the LCs of the two ZTF bands ($g$ and $r$ bands) using Lomb--Scargle method \citep{1976Ap&SS..39..447L, 1982ApJ...263..835S}. This method returns the power spectral density (PSD) over s preset range of frequencies. Since the period of $\delta$ Sct stars is very short, we set the minimum and maximum frequencies to 1 day$^{-1}$ and 1000 day$^{-1}$, respectively. We set the peak oversampling factor to 100 to ensure that each peak is adequately sampled. 

We analyzed the raw data and obtained the PSD of each $\delta$ Sct star. We recorded the frequency $f_1$ corresponding to the highest peak in the PSD and used its inverse as the first period $P_1$ (dominant period). Then based on the first period, we obtained the folded LC and used a sixth-order Fourier series (as in Eq. \ref{equation1}) to fit the LC. The difference between the maximum and minimum values of the fitted line is the amplitude ${\rm Amp}_1$ of $\delta$ Sct stars. Figure \ref{LCs1} is an example of the period determination. The upper panel shows the PSD and folded LC for the first period and the lower panel shows the PSD and folded LC for the second period.

\begin{equation}
    m=a_0+\sum a_i\cos\left(2\pi it\right)+\sum b_i\sin\left(2\pi it\right) 
    \label{equation1}
\end{equation}

To determine the second period $P_2$, we prewhitened the raw data using the fit LC of the first period $P_1$ to obtain the residual LC. Then the second period analysis was performed on the residual LC. Similarly, we used the inverse of the frequency $f_2$ corresponding to the highest peak in the PSD distribution as the second period $P_2$ and obtained the amplitude ${\rm Amp}_2$ based on the fitted LC. The top panel of Figure \ref{LCs1} shows the PSD distribution and folded LC corresponding to the first period, and the bottom panel shows the PSD distribution and folded LC corresponding to the second period. We note that the first period can be longer ($P_1>P_2$) or shorter ($P_1<P_2$) than the second period. The long period usually corresponds to F mode pulsation and the short periods correspond to 1O, 2O or 3O mode pulsation for $\delta$ Sct stars. The PSD significance of the F period may be lower than the PSD significance of the 1O, 2O or 3O period.

The reliability of the derived periods can be described by the false alarm probability (FAP). FAP is a statistical parameter indicating the probability that a random error will be considered as a periodic signal in the absence of a true periodic signal. FAP was estimated by the maximum PSD value. We excluded multi-mode $\delta$ Sct candidates with FAP values greater than 0.001. In addition, We also required that each candidate has amplitudes (${\rm Amp}_n$) larger than 0.01 mag, which is the lower limit of the accuracy of the ZTF photometry.

We set the automated procedure to continue the pre-whitening process on the residual data as long as these two conditions were met (amplitude and FAP), and searched for periods in the same way. We tried to set the initial number of periods to 16, and by calculation, we found that when more than six periods were obtained, the seventh and onwards were combinations frequencies \citep{2013AcA....63..379P,2020AcA....70..241P}. Therefore, we set the maximum number of period to six in the later processing. In this step, we obtained 11003 and 10588 candidates in $r$ and $g$ bands, respectively.

Mode classification is a complex task because most of the candidate periods do not correspond to desirable pulsation modes, so we performed the mode classification in two steps \citep{2021MNRAS.504.4039B,2022MNRAS.510.1748N}. First, we filtered each obtained frequency to exclude combination frequencies. We examined all candidate frequencies for $\delta$ Sct stars containing multiple frequencies, some of which were cases where the frequencies were added or subtracted, or equal \citep{2002A&A...394...97K,2015MNRAS.450.3015K}. For example, $f_{i} = a\ast f_{j} + b\ast f_{e}$ where $a$ and $b$ are integers, or zero. We designed an automated procedure to exclude the combination frequencies. In Table \ref{eample fre}, we show the six frequencies of some $\delta$ Sct stars obtained by our frequency determination procedure, and we can find some combination frequencies. For example, for ZTFJ174345.66+292526.7, $f_{3} = 1\ast f_{1} + 1\ast f_{2}$, $f_{4} = -1\ast f_{1} + 1\ast f_{2}$, $f_{6} = 1\ast f_{1}$, and $f_{3}$, $f_{4}$, and $f_{6}$ are the combination frequencies. Therefore, we retained $f_{1}$, $f_{2}$ and $f_{5}$, which are independent pulsation frequencies. To ensure the accuracy of the procedure of removing combination frequencies, we also calculated the ratio of combination frequencies to independent frequencies and determined that they are not in the usual sequences on the Peterson diagram of multi-mode $\delta$ Sct stars. Then, these combination frequencies were removed.

\begin{deluxetable*}{lccccccccccccc}
	\tablecaption{Example of frequencies obtained by an automatic procedure for 12 $\delta$ Sct stars.}\label{eample fre}
	\tablewidth{0.9pt}
\tablewidth{1.0pt}
	\tabletypesize{\small}
\tablehead{ZTF ID & R.A.(J2000) & decl.(J2000) & $f_1$ & $f_2$ & $f_3$ & $f_4$ & $f_5$ & $f_6$ & Passband}
	\startdata
ZTFJ174314.38+181810.6 & 265.8099413 & 18.3029559  & 18.4159   & 23.5305   & 23.6314   & 41.9464   & 35.216    & 5.1147    & $r$         \\
ZTFJ174345.66+292526.7 & 265.9402837 & 29.4240975  & 24.0333   & 30.889    & 54.9223   & 6.8557    & 22.0311   & 24.0329   & $r$         \\
ZTFJ174623.28+375715.3 & 266.5970373 & 37.9542592  & 17.4586   & 21.8274   & 17.8318   & 39.286    & 17.7307   & 4.3688    & $r$         \\
ZTFJ174643.80+285533.4 & 266.6824994 & 28.9259522  & 12.2653   & 15.7584   & 28.0238   & 3.4931    & 40.2891   & 8.7722    & $r$         \\
ZTFJ174910.76+531131.0 & 267.2948619 & 53.1919563  & 15.8378   & 15.8384   & 15.8374   & 23.4429   & 23.4434   & 23.2144   & $r$         \\
ZTFJ174923.93+152356.3 & 267.3497535 & 15.3989847  & 19.6989   & 19.9187   & 25.1866   & 20.173    & 39.8719   & 9.555     & $r$         \\
ZTFJ174945.76+510241.1 & 267.4401071 & 51.0447614  & 15.9293   & 20.4027   & 4.4734    & 36.3321   & 11.4558   & 52.2614   & $r$         \\
ZTFJ175110.34+183638.8 & 267.7931064 & 18.6107621  & 19.2241   & 24.67     & 24.8255   & 24.8368   & 5.446     & 24.8469   & $r$         \\
ZTFJ175145.87-001058.5 & 267.9411416 & -0.1829098  & 9.2821    & 12.0028   & 21.285    & 5.7261    & 17.9839   & 16.7107   & $r$         \\
ZTFJ175149.33+005540.0 & 267.955551  & 0.9277633   & 9.2299    & 9.2295    & 9.2305    & 9.229     & 9.2302    & 16.9963   & $r$         \\
ZTFJ175232.64+101403.5 & 268.1360574 & 10.2342752  & 16.6813   & 21.3826   & 4.7013    & 38.0639   & 11.98     & 54.7453   & $r$         \\
ZTFJ175354.56+273625.4 & 268.4773469 & 27.6070557  & 15.4294   & 19.6585   & 15.4299   & 35.0879   & 4.2291    & 30.8592   & $r$         \enddata
 \tablecomments{ZTF ID: Source ID; R.A. and decl.: source position (J2000); $f_1-f_6$: first to sixth frequencies, Passband: ZTF $r$ passband.}
\end{deluxetable*}

\begin{deluxetable*}{lccc}
	\tablecaption{Mode statistics of multi-mode $\delta$ Sct stars}\label{mode_count}
	\tablewidth{0.9pt}
\tablewidth{1.0pt}
	\tabletypesize{\small}
\tablehead{Mode&$N_{stars}$}
\startdata
1O/F &1753\\
2O/1O&195\\
3O/2O&96\\
3O/1O/F&38\\
2O/1O/F&44\\
3O/2O/F&9\\
3O/2O/1O&18\\
3O/1O&61\\
3O/F&16\\
2O/F&21\\
3O/2O/1O/F&3
\enddata
\end{deluxetable*}

\begin{figure*}
	\begin{center}
		\includegraphics[width=1.\linewidth]{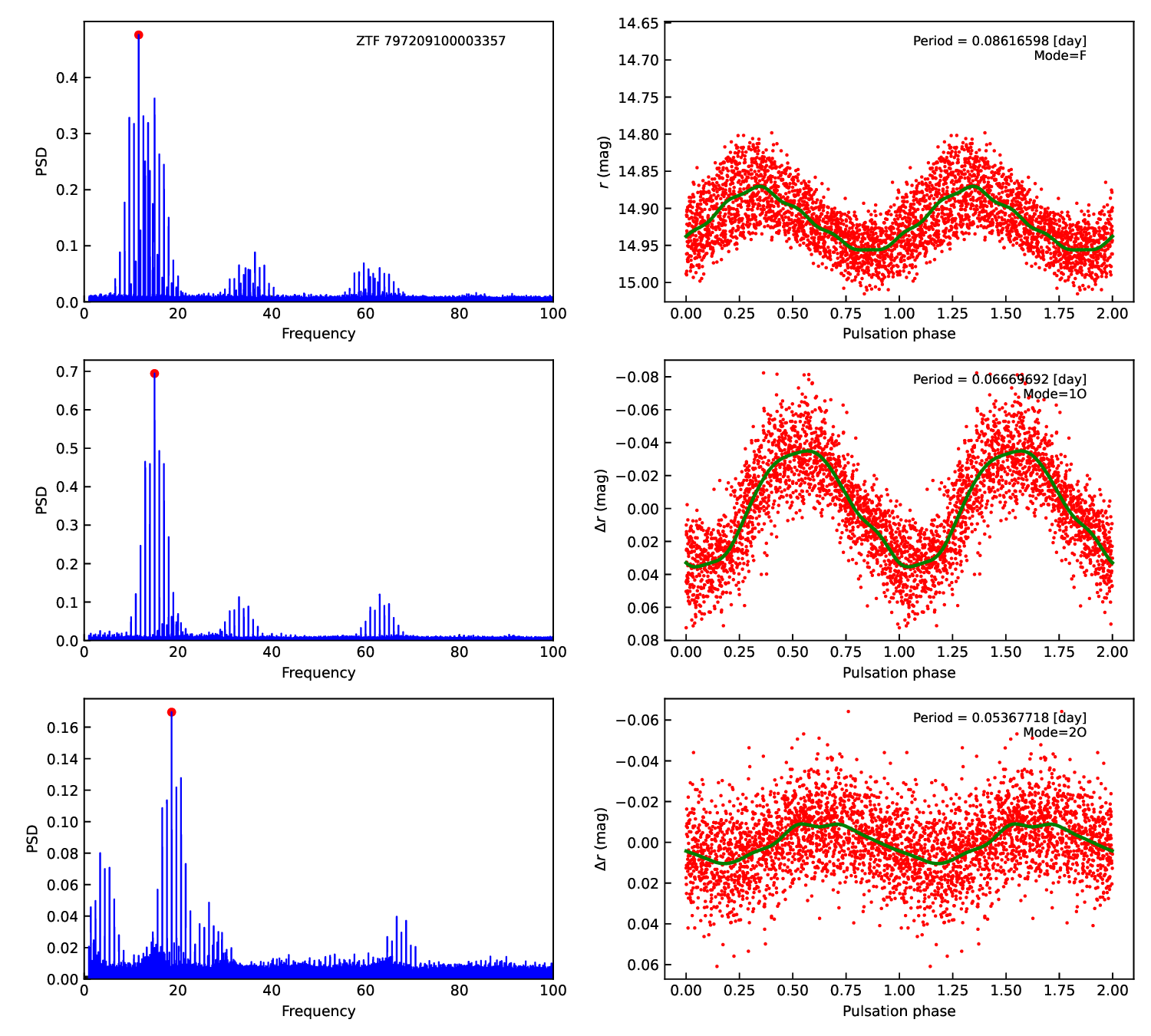}
		\caption{Example diagram of triple-mode 2O/1O/F $\delta$ Sct stars for $r$-band LC (right) and PSD (left). Top left: PSD of original data with maximum PSD marked in red. 
Top right: original LC folded by F-mode period (red dots) with best Fourier fit shown in green line. Middle left: PSD of residual LC after pre-whitening using the period corresponding to the F mode, maximum PSD marked in red. Middle right: residual LC folded by 1O mode period (red dots) with best Fourier fit shown in green line. Bottom left: PSD of residual LC after pre-whitening using the period corresponding to the 1O mode, maximum PSD marked in red. Bottom right: residual LC folded by 2O mode period (red dots) with best Fourier fit shown in green line.}
		\label{LCs2}
	\end{center}
\end{figure*}

\begin{figure*}
	\begin{center}
		\includegraphics[width=1.\linewidth]{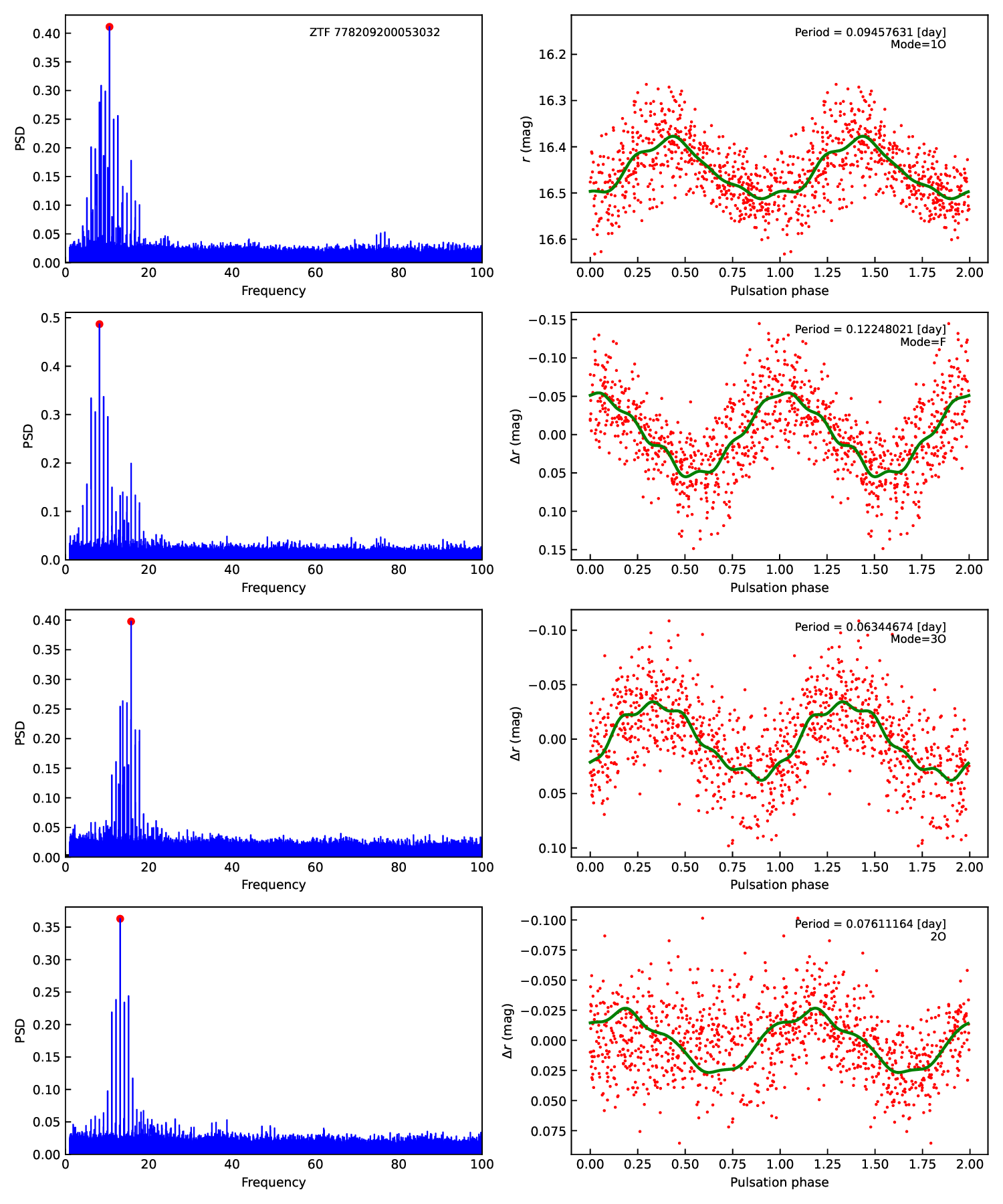}
		\caption{Example diagram of quadruple-mode 3O/2O/1O/F $\delta$ Sct stars for $r$-band LC and PSD. The left panels show the PSD plot for the corresponding mode period, with the maximum PSD marked in red. The right panels show the LC folded by corresponding mode period, with the best Fourier fit line shown in green. The panels correspond to modes 1O, F, 3O, and 2O from top to bottom.}
		\label{LCs3}
	\end{center}
\end{figure*}

We refer to the classical period ratios corresponding to each mode pair to help exclude combination frequencies \citep{2000ASPC..210....3B,2021MNRAS.504.4039B,2020AcA....70..241P,2021AcA....71..189S,2023AcA....73..105S}. We expect the largest number of real frequencies to exist simultaneously. We first screen candidates for the presence of four frequencies, namely F, 1O, 2O and 3O. If the period ratios are is in good agreement, the candidate is labeled as a quadruple-mode $\delta$ Sct candidate; otherwise, it is labeled as a triple-mode $\delta$ Sct candidate. Then, the three frequencies are analyzed to determine the corresponding modes. Some frequencies may be missing, e.g., 3O/2O/F. After selecting the triple-mode $\delta$ Sct candidates, the remaining ones are the double-mode $\delta$ Sct candidates.

After removing the combination periods, we obtained a more reliable sample, with 3890 and 3883 multi-mode $\delta$ Sct candidates in $r$ and $g$ bands, respectively. 677 $\delta$ Sct candidates exhibit triple or quadruple mode while 4460 candidates exhibit double mode.

In the second step, we perform a more strict selection of modes. Since the 3O and 2O modes have small pulsation amplitudes and the maximum PSD is more susceptible to noise, we performed a more rigorous analysis on samples classified as three or four mode, as well as on the candidates containing 2O or 3O mode.
We calculated the signal-to-noise ratio (SNR) of each maximum PSD by dividing the maximum PSD value by the average PSD value. We required that the PSD SNR of the 2O and 3O candidate periods should be more than 30 and the PSD value should be greater than 0.1 to improve the reliability of the periods. After screening, we obtained 2182 and 2172 multi-mode $\delta$ Sct candidates in the $r$ and $g$ bands, respectively. We then visually inspected the LC and PSD plots of each candidate and eliminated candidates with multiple strong PSD peaks to further ensure the reliability of each period.

Finally, we obtained 1761 and 1717 multi-mode $\delta$ Sct stars in $r$ and $g$ bands respectively. Out of these, 1224 candidates showed reliable periods in both band. If the number of radial modes identified based on the two bands is different, we suggest using the set of parameters with more modes. If the number of radial modes identified by the two bands is the same, we recommended choosing the set of parameters with the smaller second-period FAP. In total, we obtained a sample of 2254 multi-mode $\delta$ Sct stars. The count of the number of each mode is listed in Table \ref{mode_count}.

\begin{figure*}	\centering			\includegraphics[width=1.05\linewidth]{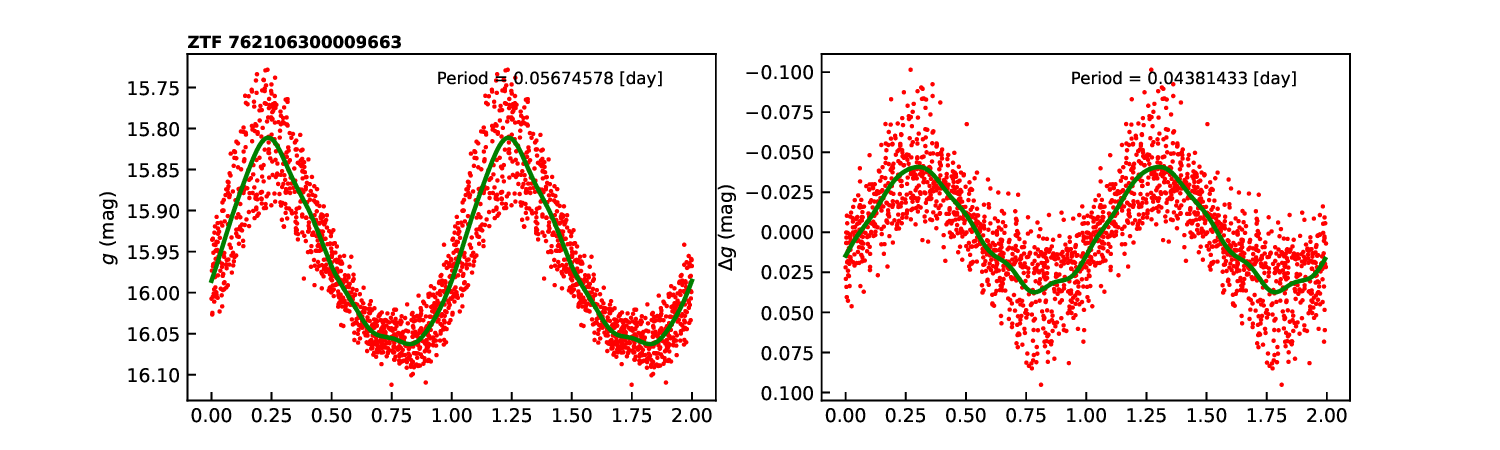}	
\includegraphics[width=1.05\linewidth]{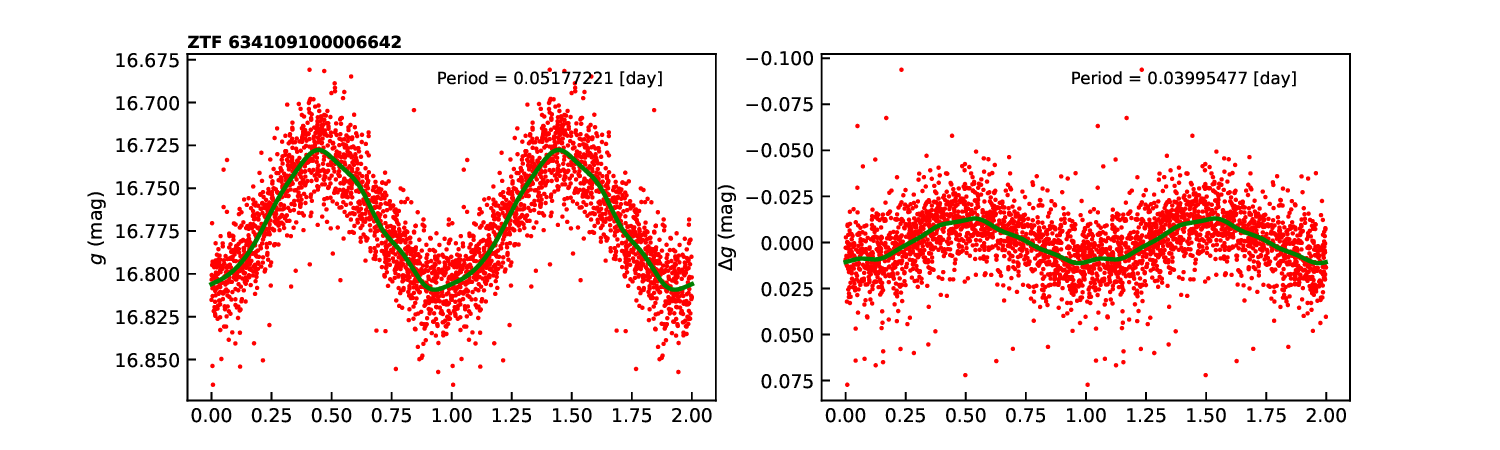}	
\includegraphics[width=1.05\linewidth]{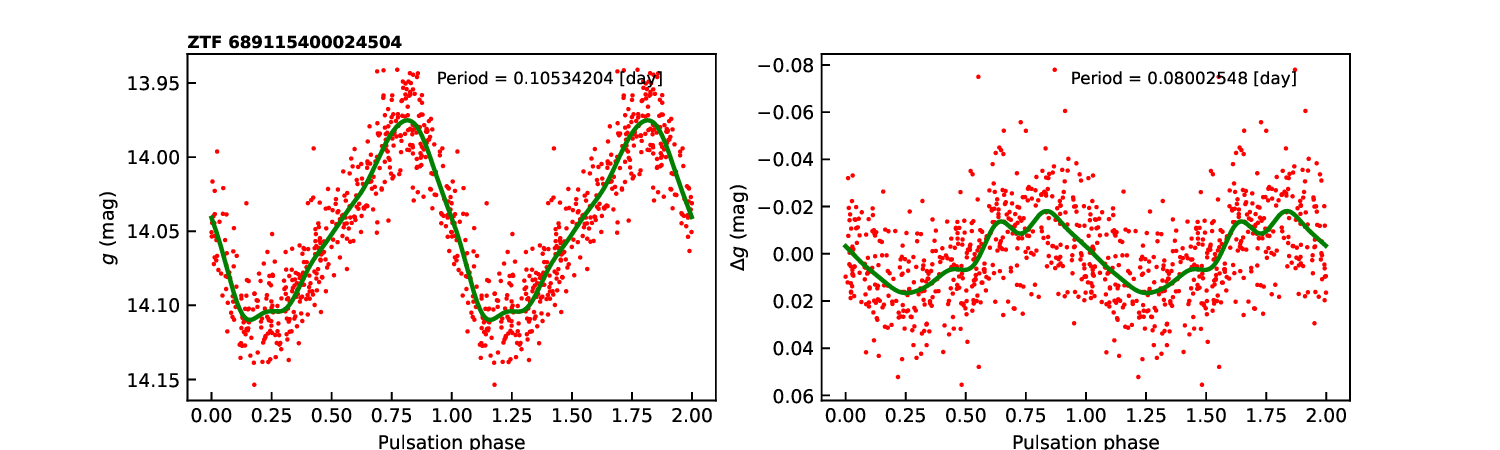}	
		\caption{Other examples of $g$-band LCs. The left panels are the original LCs folded by $P_1$, and the right panels are the residual LCs after prewhitening folded by $P_2$.}
		\label{LCs4}
\end{figure*}

Figure \ref{LCs2} illustrates an example of triple-mode $\delta$ Sct stars with 2O/1O/F mode, and Figure \ref{LCs3} illustrates a quadruple-mode $\delta$ Sct star with 3O/2O/1O/F mode. Figure \ref{LCs4} shows the characteristics of the folded LCs of 1O/F double-mode $\delta$ Sct stars. For the first period (left panels), the brightness ascending phase of most double-mode $\delta$ Sct stars takes shorter time than the brightness descending phase (top left panel). For a few double-mode $\delta$ Sct stars with low amplitudes, the brightness ascending phase takes the same time as the brightness descending phase (middle left panel) or longer (bottom left panel). For the second period (right panels), the amplitude is usually smaller than the first period. LCs for other candidates can be downloaded from the external database via \url{https://nadc.china-vo.org/res/r101344/}.

\begin{deluxetable*}{lccccccccccccc}
	\tablecaption{Periods of triple- and quadruple-mode $\delta$ Sct stars}\label{four-radial}
	\tablewidth{0.9pt}
\tablewidth{1.0pt}
	\tabletypesize{\small}
\tablehead{ZTF ID & R.A.(J2000) & decl.(J2000) & F & 1O & 2O & 3O & Passband}
	\startdata
ZTFJ002707.80+573740.7 & 6.7825489   & 57.6279757 & 0.1165398   & 0.0897679     & 0.0719473      & - -           & $g$       \\
ZTFJ003240.12+680557.2 & 8.1672032   & 68.0992404 & 0.1640398   & 0.1301509     & - -            & 0.0926695     & $r$       \\
ZTFJ011902.45+561848.4 & 19.7602403  & 56.3134413 & 0.0816329   & 0.0632725     & - -            & 0.0446231     & $g$       \\
ZTFJ015534.52+641251.5 & 28.8938852  & 64.2144173 & 0.1011103   & 0.0781726     & - -            & 0.0553498     & $g$       \\
ZTFJ022719.61+644448.5 & 36.8317323  & 64.7468217 & - -         & 0.0928653     & 0.0745411      & 0.0620484     & $g$       \\
ZTFJ024450.93+515244.9 & 41.2122211  & 51.8791557 & 0.1469574   & - -           & 0.0908331      & 0.0762280     & $g$       \\
ZTFJ041800.37+560842.4 & 64.5015703  & 56.1451017 & 0.1224803   & 0.0945763     & 0.0761113      & 0.0634466     & $g$       \\
ZTFJ051725.58+375800.0 & 79.3566291  & 37.966637  & 0.1206552   & 0.0931018     & 0.0748427      & - -           & $g$       \\
ZTFJ051955.80+330223.5 & 79.9824919  & 33.0398638 & 0.0820884   & 0.0635637     & 0.0511918      & - -           & $g$       \\
ZTFJ053635.83+180016.0 & 84.1493215  & 18.0044493 & 0.0829809   & 0.0642879     & 0.0502297      & - -           & $g$       \\
ZTFJ061454.47+082729.4 & 93.7269632  & 8.4581676  & 0.1043608   & 0.0805573     & - -            & 0.0537804     & $r$       \\
ZTFJ062314.49+132359.1 & 95.8104118  & 13.3997492 & 0.1206204   & 0.0928877     & 0.0744759      & - -           & $r$       \\
ZTFJ064324.79+212736.0 & 100.8533061 & 21.4600197 & 0.1372577   & 0.1058090     & 0.0849311      & - -           & $g$       \\
ZTFJ074121.43+514818.6 & 115.3393202 & 51.8051702 & 0.0515386   & 0.0400259     & - -            & 0.0267000     & $r$       \\
ZTFJ085208.81+475340.8 & 133.0367432 & 47.8946839 & 0.0681228   & 0.0532180     & - -            & 0.0349041     & $g$       \\
ZTFJ094705.64+320154.9 & 146.7735415 & 32.0319331 & 0.1490839   & 0.1149011     & 0.0889302      & - -           & $g$       \\
ZTFJ115725.63+461637.7 & 179.3568186 & 46.2770958 & 0.0565239   & 0.0442683     & 0.0363803      & - -           & $g$       \\
ZTFJ115938.65+283643.5 & 179.9110739 & 28.6120961 & 0.1163245   & 0.0911218     & 0.0718774      & - -           & $r$       \\
ZTFJ160627.43+611815.1 & 241.6143252 & 61.3042086 & 0.0607211   & 0.0472639     & - -            & 0.0312165     & $r$       \\
ZTFJ164334.98+314958.4 & 250.895759  & 31.832896  & 0.0802779   & 0.0618907     & 0.0482632      & 0.0402488     & $g$       \\
ZTFJ165227.35+390102.1 & 253.1139995 & 39.0172396 & 0.0758859   & 0.0587244     & 0.0473220      & - -           & $g$       \\
ZTFJ172627.88+443233.1 & 261.6162013 & 44.5425116 & 0.0517535   & 0.0402768     & - -            & 0.0256526     & $r$       \\
ZTFJ173401.09+320716.7 & 263.5045805 & 32.1213193 & 0.0819214   & 0.0615111     & - -            & 0.0416024     & $g$       \\
ZTFJ181525.21+163047.2 & 273.8550568 & 16.5131158 & 0.0916544   & 0.0707404     & 0.0549439      & - -           & $g$       \\
ZTFJ182458.41+221116.1 & 276.2434026 & 22.1878117 & 0.0773147   & 0.0605841     & - -            & 0.0407111     & $g$       \\
ZTFJ182525.08+123948.6 & 276.3545401 & 12.6634854 & 0.1023325   & 0.0789146     & 0.0107359      & - -           & $g$       \\
ZTFJ182947.45+374500.5 & 277.4477516 & 37.7501378 & 0.1165758   & 0.0902965     & - -            & 0.0635385     & $g$       \\
ZTFJ190853.44+315912.3 & 287.2227149 & 31.9867359 & 0.0942179   & 0.0728080     & 0.0583519      & - -           & $g$       \\
ZTFJ191259.09+141624.4 & 288.2462314 & 14.2734536 & - -         & 0.1107145     & 0.1339318      & 0.1113524     & $r$       \\
ZTFJ192136.47+485114.8 & 290.4020083 & 48.8541104 & 0.0795753   & 0.0642516     & - -            & 0.0422997     & $g$       \\
ZTFJ193545.79+483413.7 & 293.9408238 & 48.5704783 & 0.0954071   & 0.0748321     & 0.0603110      & - -           & $g$       \\
ZTFJ193852.49+391818.7 & 294.7187579 & 39.3052037 & 0.0666380   & 0.0515842     & - -            & 0.0349197     & $g$       \\
ZTFJ193924.92+305249.1 & 294.853881  & 30.8802924 & 0.0923304   & 0.0713013     & 0.0572283      & - -           & $r$       \\
ZTFJ194132.26+115801.8 & 295.3844368 & 11.9671693 & 0.1036147   & 0.0801730     & - -            & 0.0535791     & $g$       \\
ZTFJ194552.24+325957.6 & 296.4676933 & 32.9993304 & - -         & 0.1986043     & 0.1586355      & 0.1332206     & $r$       \\
ZTFJ194552.97+164158.8 & 296.4707254 & 16.6996697 & 0.0815354   & 0.0640002     & 0.0520505      & - -           & $g$       \\
ZTFJ194706.20+363217.0 & 296.775884  & 36.5380844 & 0.1589075   & - -           & 0.0947973      & 0.0780717     & $r$       \\
ZTFJ194926.78+200356.5 & 297.3616125 & 20.0656996 & 0.0923241   & 0.0709239     & 0.0595783      & - -           & $r$       \\
ZTFJ195026.87+445942.1 & 297.6119672 & 44.9950402 & 0.1344034   & 0.1035363     & 0.0830579      & - -           & $g$       \\
ZTFJ195805.05+400049.7 & 299.5210736 & 40.0137969 & 0.1221518   & 0.0941582     & - -            & 0.0607540     & $r$       \\
ZTFJ200045.52+452712.4 & 300.1897125 & 45.45344   & 0.0614813   & 0.0477953     & 0.0373422      & - -           & $r$       \\
ZTFJ200509.43+415657.6 & 301.2893281 & 41.9493179 & 0.1283890   & - -           & 0.0821899      & 0.0689944     & $g$       \\
ZTFJ201154.37+235213.4 & 302.9765662 & 23.8703791 & 0.0944245   & 0.0755396     & 0.0609555      & - -           & $r$       \\
ZTFJ202307.41+113823.2 & 305.7809105 & 11.6397825 & 0.0667728   & 0.0515535     & 0.0415349      & - -           & $g$       \\
\multicolumn{1}{c}{...} & \multicolumn{1}{c}{...} & \multicolumn{1}{c}{...} & \multicolumn{1}{c}{...} & \multicolumn{1}{c}{...} & \multicolumn{1}{c}{...} & \multicolumn{1}{c}{...} & \multicolumn{1}{c}{...}     \\ZTFJ173856.54+270327.3 & 264.735597  & 27.0575901 & 0.0746851   & 0.0566629     & - -            & 0.0373380      & $g$       \\
ZTFJ190503.39+162956.8 & 286.2641608 & 16.4991235 & 0.0814394   & 0.0629045     & 0.0497054      & 0.0405029     & $g$       \\
ZTFJ192406.37+264808.3 & 291.026599  & 26.8023039 & 0.1962647   & 0.1499658     & 0.1201918      & - -           & $g$    
\enddata
 \tablecomments{ZTF ID: source ID; R.A. and decl.: source position (J2000); F: fundamental period, 1O: first-overtone period, 2O: second-overtone period, 3O: third-overtone period, Passband: ZTF passband. "- -" indicates that this mode is not detected.\\ (This table is available in its entirety in machine-readable form.)}
\end{deluxetable*}

\begin{deluxetable*}{lccccccccc}
	\tablecaption{Multi-mode $\delta$ Sct stars in ZTF DR20}
	\tablewidth{1.2pt}
\tablewidth{1.0pt}
	\tabletypesize{\small}
	\tablehead{ZTF ID & R.A. (J2000) & decl. (J2000) & \multicolumn{1}{c}{...} & $P_{F}^{r}$ & PSD$_{F}^{r}$ & $\log$ FAP$_{F}^{r}$ &  Amp$_{F}^{r}$ & ...& Amp$_{3O}^{g}$\\ & (deg) & (deg)& \multicolumn{1}{c}{...} & (days) &  &  & (mags)& &(mags) }
 \label{parameter}
	\startdata
ZTFJ000013.16+514411.1 & 0.0548631   & 51.7364071  & ...                 & 0.0698843      & 0.6049              & -137.898            & 0.046               & ...                & - -                   \\
ZTFJ000049.07+551428.0 & 0.2045118   & 55.2411286  & ...                 & 0.0766785      & 0.6815              & -168.968            & 0.245               & ...                & - -                   \\
ZTFJ000114.74+534305.4 & 0.3114546   & 53.7181745  & ...                 & 0.0789014      & 0.5882              & -132.138            & 0.126               & ...                & - -                   \\
ZTFJ000123.15+681742.6 & 0.3464857   & 68.295184   & ...                 & 0.0825984      & 0.5618              & -173.124            & 0.338               & ...                &                       \\
ZTFJ000316.54+625516.9 & 0.8189658   & 62.9213692  & ...                 & 0.1768517      & 0.0883              & -8.366            & 0.016               & ...                & - -                   \\
ZTFJ000406.53+633118.4 & 1.0272561   & 63.5217831  & ...                 &                &                     &                     &                     & ...                & - -                   \\
ZTFJ000409.21+570516.3 & 1.0384164   & 57.0878665  & ...                 & - -            & - -                 & - -                 & - -                 & ...                &                       \\
ZTFJ000441.51+572904.8 & 1.1729595   & 57.4846612  & ...                 & 0.0762576      & 0.8766              & -320.000                & 0.312               & ...                & - -                   \\
ZTFJ000442.98+534047.8 & 1.1791283   & 53.6799309  & ...                 &                &                     &                     &                     & ...                & - -                   \\
ZTFJ000508.12+571258.9 & 1.2838952   & 57.2163783  & ...                 & 0.0438734      & 0.7370              & -320.000                & 0.191               & ...                & - -                   \\
ZTFJ000514.81+323303.1 & 1.3117522   & 32.5508635  & ...                 &                &                     &                     &                     & ...                & - -                   \\
ZTFJ000536.90+612746.7 & 1.4037999   & 61.4629866  & ...                 &                &                     &                     &                     & ...                & 0.021                 \\
ZTFJ000634.49+611228.6 & 1.6437056   & 61.2079585  & ...                 & 0.0806640      & 0.7643              & -204.250            & 0.113               & ...                & - -                   \\
ZTFJ000639.44+531344.0 & 1.6643797   & 53.2288917  & ...                 & - -            & - -                 & - -                 & - -                 & ...                & 0.039                 \\
ZTFJ000707.20+611036.2 & 1.7799918   & 61.1767359  & ...                 & 0.0873930      & 0.5830              & -121.751            & 0.170               & ...                &                       \\
ZTFJ000850.77+183109.8 & 2.2115708   & 18.5193646  & ...                 & 0.0752332      & 0.1866              & -22.650            & 0.028               & ...                & - -                   \\
ZTFJ000933.50+562144.3 & 2.3896312   & 56.3623363  & ...                 & 0.0836583      & 0.6713              & -286.332            & 0.194               & ...                & - -                   \\
ZTFJ001020.14+493126.7 & 2.5839575   & 49.5240795  & ...                 &                &                     &                     &                     & ...                & - -                   \\
ZTFJ001023.45+603338.9 & 2.5977259   & 60.5608094  & ...                 & 0.1718441      & 0.7759              & -199.861            & 0.082               & ...                &                       \\
ZTFJ001032.32+322348.0 & 2.6346907   & 32.3966982  & ...                 &                &                     &                     &                     & ...                & - -                   \\
ZTFJ001048.18+561927.4 & 2.7007918   & 56.3242795  & ...                 & - -            & - -                 & - -                 & - -                 & ...                & 0.027                 \\
ZTFJ001052.49+563118.6 & 2.7187621   & 56.5218396  & ...                 & 0.1539481      & 0.6317              & -241.448            & 0.183               & ...                &                       \\
ZTFJ001133.52+562739.7 & 2.8897211   & 56.4610453  & ...                 &                &                     &                     &                     & ...                & 0.043                 \\
ZTFJ001350.50+500600.3 & 3.460446    & 50.1001078  & ...                 &                &                     &                     &                     & ...                & - -                   \\
ZTFJ001424.71+532052.6 & 3.6030072   & 53.3479601  & ...                 & 0.0859501      & 0.9102              & -320.000                & 0.252               & ...                & - -                   \\
ZTFJ001452.28+595603.4 & 3.7178555   & 59.9342835  & ...                 & - -            & - -                 & - -                 & - -                 & ...                & 0.076                 \\
ZTFJ001454.08+705311.7 & 3.725387    & 70.8866208  & ...                 & 0.0787709      & 0.8523              & -221.559            & 0.212               & ...                &                       \\
ZTFJ001508.70+434022.1 & 3.7862786   & 43.672824   & ...                 & 0.0707618      & 0.6352              & -170.469            & 0.396               & ...                &                       \\
ZTFJ001616.57+122320.5 & 4.0690587   & 12.3890231  & ...                 & 0.0490582      & 0.8905              & -241.631            & 0.285               & ...                & - -                   \\
ZTFJ001737.44+613711.4 & 4.4060306   & 61.6198361  & ...                 & - -            & - -                 & - -                 & - -                 & ...                &                       \\
ZTFJ001825.64+670205.9 & 4.6068601   & 67.034985   & ...                 & 0.0733460      & 0.7245              & -161.080            & 0.135               & ...                & - -                   \\
ZTFJ001958.76+460858.8 & 4.9948454   & 46.1496626  & ...                 & 0.0591786      & 0.8123              & -88.713            & 0.329               & ...                &                       \\
ZTFJ002132.58+182817.4 & 5.3857876   & 18.4715083  & ...                 & 0.0653385      & 0.2477              & -35.938            & 0.112               & ...                & - -                   \\
ZTFJ002331.41+713953.2 & 5.8809224   & 71.6647767  & ...                 & 0.0820087      & 0.7447              & -165.862            & 0.127               & ...                & - -                   \\
ZTFJ002345.17+644554.8 & 5.9382365   & 64.7652377  & ...                 & 0.1097914      & 0.5831              & -111.574            & 0.097               & ...                & - -                   \\
ZTFJ002507.56+632931.9 & 6.2815526   & 63.4922072  & ...                 &                &                     &                     &                     & ...                & - -                   \\
ZTFJ002622.93+562517.2 & 6.5955714   & 56.4214408  & ...                 & 0.0723140      & 0.7183              & -182.121            & 0.213               & ...                & - -                   \\
ZTFJ002707.80+573740.7 & 6.7825467   & 57.6279743  & ...                 & 0.1165400      & 0.4041              & -72.155             & 0.052               & ...                & - -                   \\
ZTFJ002725.50+371018.8 & 6.8562575   & 37.1718835  & ...                 & 0.0527894      & 0.6876              & -200.862            & 0.182               & ...                & - -                   \\
ZTFJ002843.66+642637.2 & 7.1819293   & 64.4436954  & ...                 & 0.0807416      & 0.2081              & -27.366             & 0.047               & ...                &                       \\
ZTFJ002928.84+541842.6 & 7.3702144   & 54.311838   & ...                 & 0.0620472      & 0.8251              & -271.473            & 0.236               & ...                & - -                   \\
\multicolumn{1}{c}{...} & \multicolumn{1}{c}{...}  & \multicolumn{1}{c}{...} & \multicolumn{1}{c}{...} & \multicolumn{1}{c}{...}     & \multicolumn{1}{c}{...}             & \multicolumn{1}{c}{...}             & \multicolumn{1}{c}{...}           & \multicolumn{1}{c}{...} & \multicolumn{1}{c}{...}                 \\
ZTFJ005008.94+411040.4 & 12.5372821  & 41.177881   & ...                 & 0.0554433      & 0.5507              & -165.096            & 0.180               & ...                & - -                                
\enddata
	\tablecomments{$Mag_{F}^{r}$, $P_{F}^{r}$, PSD$_{F}^{r}$, $\log$ FAP$_{F}^{r}$, Amp$_{F}^{r}$: mean magnitude, period, maximum power spectral density, false alarm probability and amplitude corresponding to F mode in ZTF $r$ band. $P_{1O}^{r}$, PSD$_{1O}^{r}$, $\log$ FAP$_{1O}^{r}$, Amp$_{1O}^{r}$: period, maximum power spectral density, false alarm probability and amplitude corresponding to 1O mode in ZTF $r$ band. $P_{2O}^{r}$, PSD$_{2O}^{r}$, $\log$ FAP$_{2O}^{r}$, Amp$_{2O}^{r}$: period, maximum power spectral density, false alarm probability and amplitude corresponding to 2O mode in ZTF $r$ band. $P_{3O}^{r}$, PSD$_{3O}^{r}$, $\log$ FAP$_{3O}^{r}$, Amp$_{3O}^{r}$: period, maximum power spectral density, false alarm probability and amplitude corresponding to 3O mode in ZTF $r$ band. \\"- -" indicates that this mode is not detected.\\
(This table is available in its entirety in machine-readable form.)}
\end{deluxetable*}

\section{Results}\label{sec:result}

\begin{deluxetable*}{lcccccccc}
	\tablecaption{Comparison of the first periods and second periods of multi-mode $\delta$ Sct stars determined by our and the OGLE catalog}
	\tablewidth{1.0pt}
	\tabletypesize{\large}
	\tablehead{ZTF ID & R.A.(J2000) & decl.(J2000) & OGLE $P_1$ & ZTF $P_1$ & OGLE $P_2$ & ZTF $P_2$ \\ & (deg) & (deg) & (days) & (days) & (days) & (days) }
 \label{matched}
	\startdata
ZTFJ054655.19+155908.3 & 86.72996  & 15.98564  & 0.08503635 & 0.0850360 & 0.0659124  & 0.0659121 \\
ZTFJ060440.42+083432.3 & 91.16845  & 8.57566   & 0.07191587 & 0.0719158 & 0.04854491 & 0.0485444 \\
ZTFJ061355.52+034318.0 & 93.48134  & 3.72168   & 0.08596043 & 0.0859603 & 0.06677114 & 0.0667712 \\
ZTFJ062413.66+050418.0 & 96.05695  & 5.07169   & 0.15025444 & 0.1502542 & 0.1155531  & 0.1155533 \\
ZTFJ062547.82+191921.0 & 96.44928  & 19.3225   & 0.1413627  & 0.1413625 & 0.10852826 & 0.1085282 \\
ZTFJ062552.25-005202.6 & 96.46774  & -0.86739  & 0.08598725 & 0.0859880 & 0.06667738 & 0.0666785 \\
ZTFJ062824.41+035852.5 & 97.10174  & 3.98126   & 0.14158798 & 0.1415881 & 0.12732041 & 0.112906$^b$  \\
ZTFJ063629.09-051433.9 & 99.12124  & -5.24276  & 0.138682   & 0.1386829 & 0.18032694 & 0.1803272 \\
ZTFJ063753.01-033949.7 & 99.4709   & -3.66383  & 0.13108672 & 0.1310867 & 0.10104135 & 0.1010403 \\
ZTFJ064304.96+080714.7 & 100.77068 & 8.12076   & 0.13705798 & 0.1370576 & 0.10903951 & 0.1090399 \\
ZTFJ064320.31+060135.4 & 100.83463 & 6.02651   & 0.1174762  & 0.1174762 & 0.09772504 & 0.0977247 \\
ZTFJ064717.99+061003.3 & 101.82499 & 6.16759   & 0.12671519 & 0.1267154 & 0.14258489 & 0.1030856$^b$ \\
ZTFJ065247.18-090306.6 & 103.19659 & -9.05185  & 0.09385503 & 0.0938550 & 0.07268911 & 0.0726892 \\
ZTFJ065350.50-130017.1 & 103.46042 & -13.00477 & 0.14432489 & 0.1443258 & 0.12594376 & 0.1118607$^b$ \\
ZTFJ065407.53+050434.3 & 103.53141 & 5.07622   & 0.10114655 & 0.1011467 & 0.07807166 & 0.0780718 \\
ZTFJ065704.42-083331.2 & 104.26843 & -8.55868  & 0.12992842 & 0.1299283 & 0.09998667 & 0.0999870 \\
ZTFJ065818.31-021908.2 & 104.57632 & -2.31895  & 0.09357905 & 0.0935790 & 0.07247291 & 0.0724733 \\
ZTFJ184116.17-115906.6 & 280.31739 & -11.98518 & 0.06057093 & 0.0605708 & 0.04689278 & 0.0468926 \\
ZTFJ184333.95+044143.6 & 280.89147 & 4.69546   & 0.04798205 & 0.0619148 & 0.06191417 & 0.0479824$^a$ \\
ZTFJ184532.70+092424.2 & 281.38628 & 9.40673   & 0.06176577 & 0.0617659 & 0.04775663 & 0.0477563 \\
ZTFJ184607.72+124928.0 & 281.53219 & 12.82445  & 0.06448798 & 0.0644882 & 0.04999730 & 0.0499978 \\
ZTFJ184807.47-054828.5 & 282.03113 & -5.80794  & 0.08144335 & 0.0814431 & 0.06462169 & 0.0646215 \\
ZTFJ184837.05+123057.7 & 282.15439 & 12.51604  & 0.10062764 & 0.0776992 & 0.07769838 & 0.1006291$^a$ \\
ZTFJ185121.24+004235.4 & 282.83851 & 0.70986   & 0.09587564 & 0.0958755 & 0.07404729 & 0.0740473 \\
ZTFJ185336.75-034220.8 & 283.40313 & -3.70578  & 0.07366549 & 0.0736642 & 0.09469556 & 0.0946936 \\
ZTFJ185452.44-052428.0 & 283.71852 & -5.40778  & 0.18920317 & 0.1892029 & 0.14466831 & 0.1446693 \\
ZTFJ190543.74-052401.3 & 286.43229 & -5.40037  & 0.07401666 & 0.0740168 & 0.05715781 & 0.0571581 \\
ZTFJ190620.59+145937.3 & 286.58582 & 14.99372  & 0.08148242 & 0.0814825 & 0.06303693 & 0.0630369 \\
ZTFJ190704.49-000737.3 & 286.76875 & -0.12705  & 0.07215581 & 0.0721557 & 0.05586513 & 0.0558649 \\
\multicolumn{1}{c}{...} & \multicolumn{1}{c}{...}  & \multicolumn{1}{c}{...} & \multicolumn{1}{c}{...}     & \multicolumn{1}{c}{...}             & \multicolumn{1}{c}{...}             & \multicolumn{1}{c}{...}   \\
ZTFJ190850.13+010116.0 & 287.20889 & 1.02112   & 0.08203788 & 0.0820377 & 0.06394150 & 0.0639415 \\
ZTFJ191229.13-030916.9 & 288.1214  & -3.15472  & 0.08440821 & 0.0844081 & 0.07011598 & 0.0655097$^b$\\
ZTFJ191306.40+210539.4 & 288.27669 & 21.09428  & 0.14697661 & 0.1469794 & 0.11429987 & 0.1142993 
	\enddata
	\tablecomments{OGLE $P_1$, ZTF $P_1$: first periods determined by OGLE and ZTF DR20 data; OGLE $P_2$, ZTF $P_2$: second periods determined by OGLE and ZTF DR20 data.\\
	$^a$ ZTF $P_1$ equals to OGLE $P_2$ and ZTF $P_2$ equals to OGLE $P_1$. \\
	$^b$ objects with similar $P_1$ but different $P_2$.\\
(This table is available in its entirety in machine-readable form.)}
\end{deluxetable*}

In total, we find 2254 multi-mode $\delta$ Sct, with 2142 double-mode $\delta$ Sct, 109 triple-mode $\delta$ Sct, and 3 quadruple-mode $\delta$ Sct. Table \ref{four-radial} shows the periods of $\delta$ Sct with three or four radial modes, with columns 4 through 7 showing the periods of the F, 1O, 2O, and 3O modes, respectively. The last column records the band used to obtain these periods. 

Table \ref{parameter} lists the parameters of 2254 multi-mode $\delta$ Sct stars. These parameters include position, period of each mode and the corresponding PSD, FAP and amplitude, with a null value indicating that the mode has not been detected. The online version of this table also includes the individual parameters measured in the $g$ and $r$ bands.

The F mode has a maximum period of 0.29088 days, a minimum period of 0.04 days, and an average value of 0.08171 days. For the 1O mode, the maximum, minimum and average periods are 0.22346 days, 0.03217 days, and 0.06765 days, respectively. The 2O mode has maximum and minimum periods of 0.18946 days and 0.03046 days, and an average value of 0.08842 days. The 3O mode has maximum and minimum periods of 0.15931 days and 0.02367 days, and an average value of 0.07118 days. The more discussion on the statistical properties about 1O/F of these parameters is in Section \ref{sec:statistics}. 

To better evaluate the confidence of our sample, we cross-matched it with the OGLE catalog which includes 4023 and 1838 double-mode $\delta$ Sct stars in the Galactic bulge and the Galactic southern disk \citep{2020AcA....70..241P}. We found that 73 multi-mode $\delta$ Sct stars were included in OGLE's catalog (see Table \ref{matched}). The majority of them are in 1O/F mode, and a few are in other modes. By comparing the first periods $P_1$ determined by ZTF and OGLE data, we found that 71 $\delta$ Sct stars are consistent, if we considered that the period difference is not greater than 0.00001. For the other two $\delta$ Sct stars, our two periods are just the opposite of the two periods determined by OGLE (see Table \ref{matched}, labeled $^a$). That is, ZTF $P_1$ = OGLE $P_2$ and ZTF $P_2$ = OGLE $P_1$. For both of them, the amplitudes and maximum PSDs of their two periods are relatively close to each other, and the strongest period may alternate in different ephemerides (${\rm HJD}_{\rm OGLE}<2458762$, ${\rm HJD}_{\rm ZTF}>2458205$). For the second period $P_2$, the period differences of 69 $\delta$ Sct stars are less than 0.0001. Four $\delta$ Sct stars show differences on the $P_2$ (see Table \ref{matched}, labeled $^b$). For these four, OGLE's period ratio is significantly deviated from 0.77, and their second period is likely the other mode or an aliased period. 

Overall, the average proportion of our two periods that agree with the OGLE periods is 95.8\%. Considering that the second period of $\delta$ Sct stars is more difficult to identify than that of RR Lyrae, this proportion shows that our sample of multi-mode $\delta$ Sct stars is reliable.

\section{Discussion}\label{sec:statistics}

\begin{figure*}
	\begin{center}
\includegraphics[width=0.9\linewidth]{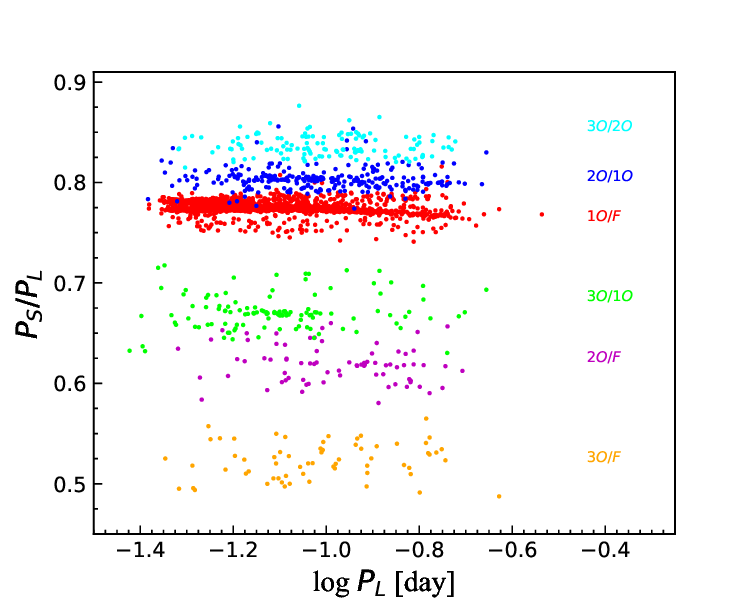}
		\caption{Petersen diagram of mutlti-mode $\delta$ Sct stars. The period ratio (y axis) is the short period ($P_S$) divided by the long period ($P_L$) . }
		\label{petersen}
	\end{center}
\end{figure*}

The Petersen diagram of multi-mode $\delta$ Sct stars is shown in Figure \ref{petersen}. $P_S/P_L$ represents the ratio of the shorter and longer periods, and $\log (P_L) $ is the logarithm of the longer periods. Different colored dots show different modes: red dots indicate the 1O/F mode, blue dots indicate the 1O/2O mode, cyan dots indicate the 3O/2O mode, green dots indicate the 3O/1O mode, magenta dots indicate the 2O/F mode, and orange dots indicate the 3O/F mode, respectively. It is worth noting that there may be mixing between different modes on the Peterson diagram. For a $\delta$ Sct star, when only two periods are detected, we classify them by strict ranges of period ratios. At this point there will be a small number of $\delta$ Sct stars that are misclassified, such as the modes of 1O/F and 2O/1O. When there are three or four periods detected, we can classify the modes by multiple sets of period ratios, and the classification accuracy will be higher.

\begin{figure*}
	\begin{center}
	\includegraphics[width=1.0\linewidth]{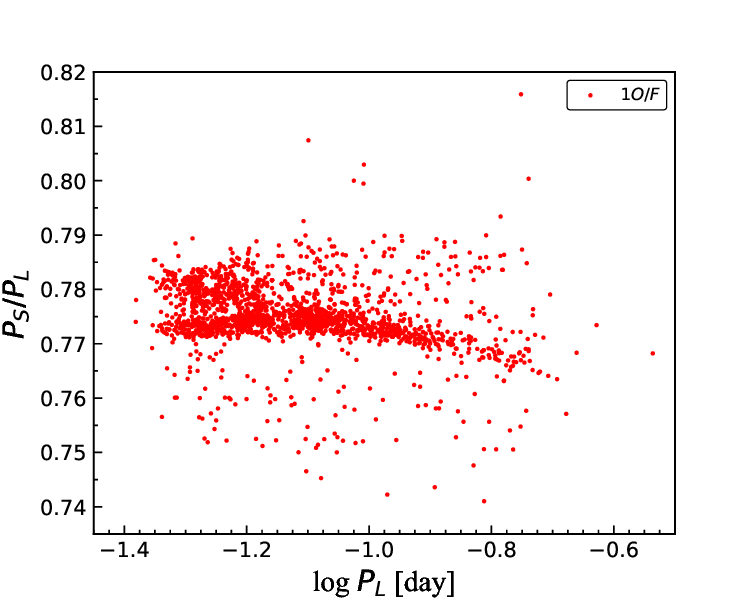}
  \caption{Petersen diagram of 1O/F double-mode $\delta$ Sct stars. The period ratio (y axis) is the 1O mode period divided by the F mode period. }
		\label{petersen1O}
	\end{center}
	\end{figure*}

Multi-mode $\delta$ Sct stars can be used to obtain stellar masses and luminosities, and can also be used as a good distance tracer. Here, we focus on discussing the statistical properties of 1O/F mode $\delta$ Sct stars, which account for the majority of multi-period $\delta$ Sct stars.
We draw their Petersen diagram separately in Figure \ref{petersen1O}. On this diagram, we can see a distinct sequence with period ratios ranging from 0.760 to 0.775 \citep{2000ASPC..210....3B,2000PASP..112.1096M} and a clump with a period ratio around 0.78. This clump was also found by \cite{2021AcA....71..189S}.  According to \citet{2000ASPC..210....3B}, the sequence corresponds to Population I $\delta$ Sct stars, while the clump is likely formed by Population II SX Phe stars. We found that the amplitude and amplitude ratio distribution of the clump and the sequence is similar.

\begin{figure*}
	\begin{center}
	\includegraphics[width=1.0\linewidth]{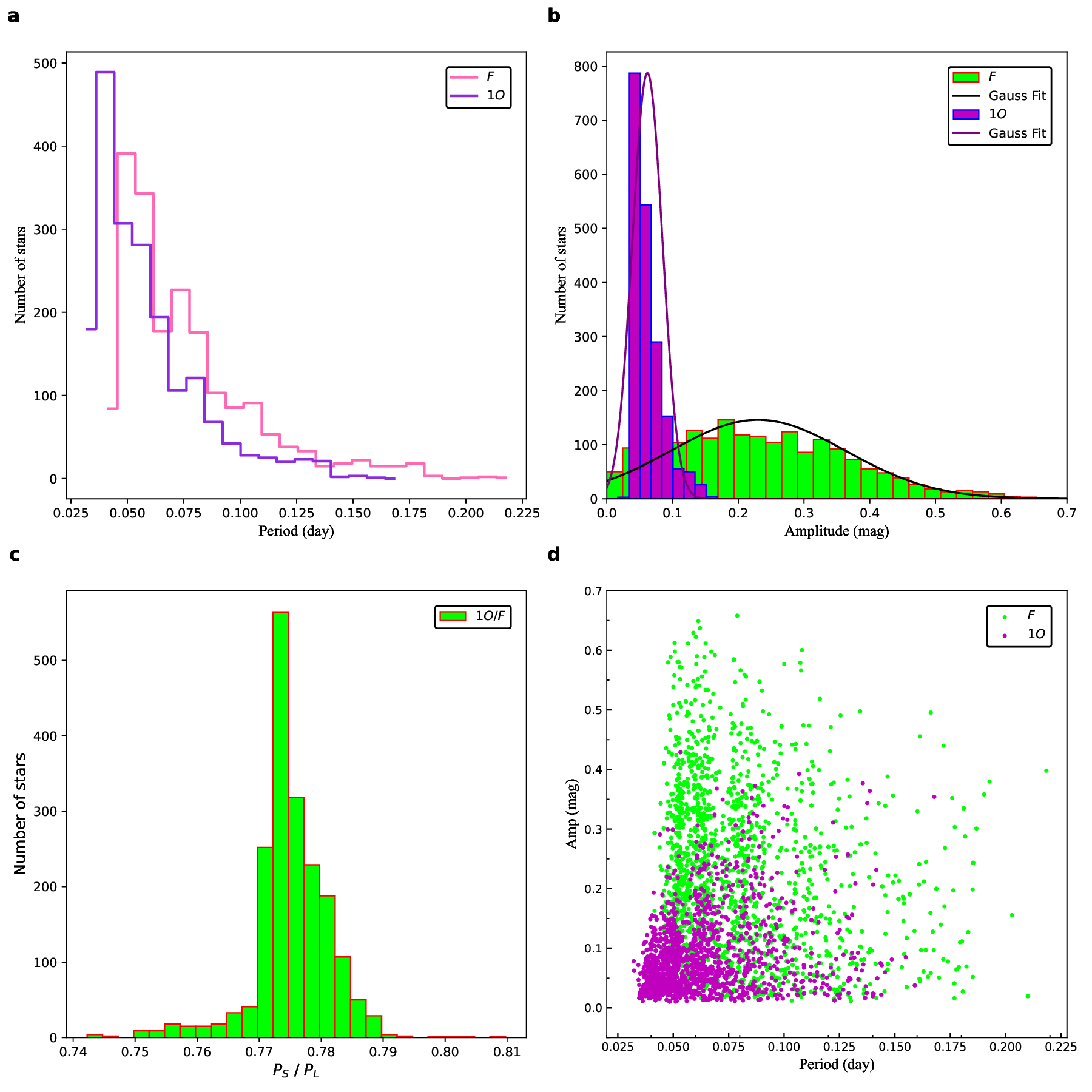}
		\caption{Statistical properties of the 1O/F double-mode $\delta$ Sct stars. a: histograms of the F mode period and 1O mode period; b: histograms of the amplitude corresponding to F and 1O mode, the two solid lines are the Gaussian fits; c: histograms of the period ratio; d: distribution of ${\rm Amp}$ vs. period.}
		\label{statics}
	\end{center}
	\end{figure*}

Figure \ref{statics}a shows the histograms of the F (green) and 1O (magenta) mode period for 1O/F double-mode $\delta$ Sct stars. We can see that the distribution of the two periods is similar, with most of them being less than 0.6 days. Figure \ref{statics}b shows the histograms of the amplitude, and we performed a Gaussian fit to the histograms (two solid curves). For F and 1O modes, $\mu$ are 0.230 mag and 0.062 mag, while $\sigma$ are 0.133 mag and 0.023 mag, respectively. For $\delta$ Sct stars with 1O mode, their amplitudes are smaller than 0.2 mag, which is significantly smaller than that of the F mode. Figure \ref{statics}c is the histogram of the period ratios. It is clear that the period ratios of $\delta$ Sct stars are almost concentrated between 0.75 and 0.79 with a peak at about 0.774. Figure \ref{statics}d shows the distribution of amplitude vs. period, the amplitude of 1O mode is smaller than that of F mode.


\begin{figure*}
	\centering				
     \includegraphics[width=1.05\linewidth]{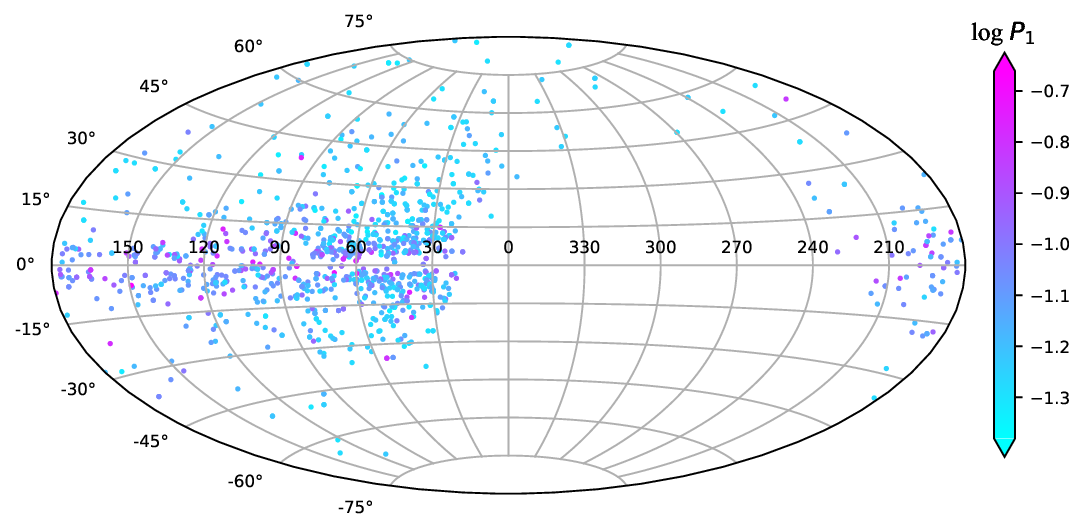}\\{ (a) }	
	   \hspace{1cm}
	\includegraphics[width=1.05\linewidth]{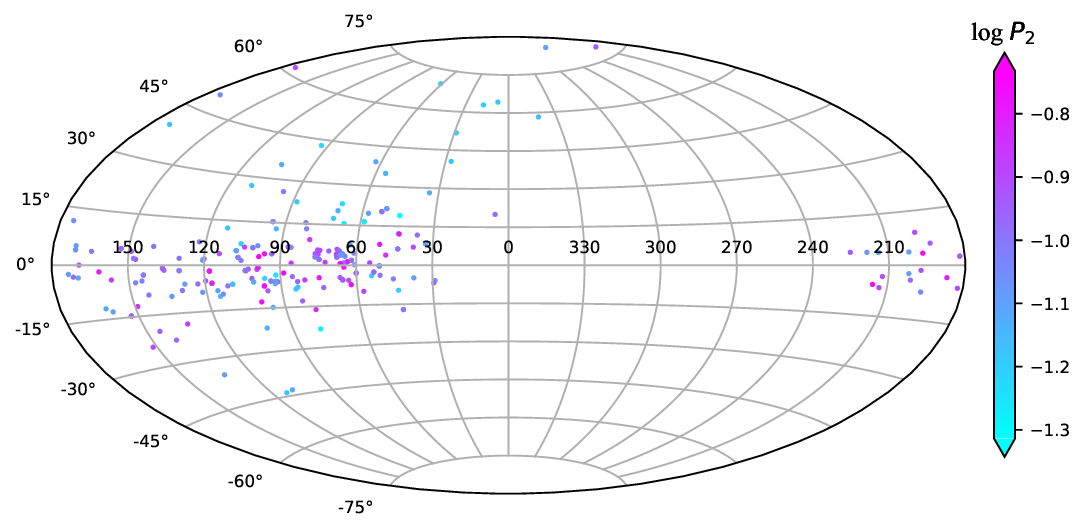}\\{ (b) }
		\caption{All-sky distribution of 1O/F double-mode $\delta$ Sct stars in Galactic coordinates. Top panel shows the sample with $P_1$ $>$ $P_2$,
while bottom panel shows the sample with $P_1$ $<$ $P_2$. The colors indicates the logarithm of the longer period $P_L$.}
		\label{allsky}
	\end{figure*}

Figure \ref{allsky} shows the all-sky distribution of these 1O/F double-mode $\delta$ Sct stars in Galactic coordinates. We divided them into two samples: the first period is F mode and the second period is F mode. These two samples represent F-mode PSDs that are greater or less than 1O-mode PSDs, respectively. The colors show the logarithm of the longer period $P_L$. We find that the period increases from the high latitude to the Galactic plane for the two samples, which was also found by \cite{2021AcA....71..189S}. This trend of period distribution is caused by the difference in metallicities, and the mean pulsation period of the metal-poor $\delta$ Sct stars is shorter than those of the metal-rich $\delta$ Sct stars \citep{2020MNRAS.493.4186J}.\\

\section{conclusion}\label{sec:conclusion}
We designed procedures to obtain periods and exclude combination periods based on $\delta$ Sct stars' LCs and residual LCs. Then, we obtained a sample of 2254 multi-mode $\delta$ Sct stars based on ZTF DR20, of which 2181 are newly discovered. This is the largest sample in the northern sky. For 1761 and 1717 multi-mode $\delta$ Sct stars, we determined the $g$- and $r$-band periods, pulsation mode, amplitudes, and the FAPs of the periods, respectively. 2142 objects are double-mode $\delta$ Sct stars, while 109 objects are triple-mode $\delta$ Sct and 3 are quadruple-mode $\delta$ Sct. 

Different multi-mode $\delta$ Sct stars have different distributions on the Petersen diagram, among which we focused on the 1O/F double-modal $\delta$ Sct stars. They are presented as a sequence and a clump at short period end on the Petersen diagram.
In Galactic coordinates, the period of 1O/F double-mode $\delta$ Sct stars at high latitudes is shorter than that on the disk, which is caused by different metallicities. Short-period double-mode $\delta$ Sct stars have lower metallicities. In the future, this sample of 1O/F double-mode $\delta$ Sct stars will help to better understand the internal structure and evolutionary history of $\delta$ Sct stars, as well as to constrain their masses. Double-mode $\delta$ Sct stars can be used as standard candles to measure distances to clusters and dwarf galaxies, and can also be used to study the structure and evolution of the Milky Way.

\section{Acknowledgements}
We thank the anonymous reviewer for the useful comments. This work was supported by the National Natural Science Foundation of China (NSFC) through grants 12173047, 12322306, 12003046, 12373028, 12233009, 12133002, 11903045, 12173028. X. Chen and S. Wang acknowledge support from the Youth Innovation Promotion Association of the Chinese Academy of Sciences (No. 2022055 and 2023065). We also thanked the support from the National Key Research and development Program of China, grants 2022YFF0503404.
This work is based on observations obtained with the Samuel Oschin Telescope 48-inch and the 60-inch Telescope at the Palomar Observatory as part of the Zwicky Transient Facility project. ZTF is supported by the National Science Foundation under Grants No. AST-1440341 and AST-2034437 and a collaboration including current partners Caltech, IPAC, the Weizmann Institute for Science, the Oskar Klein Center at Stockholm University, the University of Maryland, Deutsches Elektronen-Synchrotron and Humboldt University, the TANGO Consortium of Taiwan, the University of Wisconsin at Milwaukee, Trinity College Dublin, Lawrence Livermore National Laboratories, IN2P3, University of Warwick, Ruhr University Bochum, Northwestern University and former partners the University of Washington, Los Alamos National Laboratories, and Lawrence Berkeley National Laboratories. Operations are conducted by COO, IPAC, and UW.

\bibliographystyle{aasjournal} 
\bibliography{MM_DSCT} 


\end{document}